\documentclass[superscriptaddress,twocolumn,pra,nofootinbib]{revtex4-2}
\usepackage{graphicx}
\usepackage{calc}
\usepackage{amsmath}
\usepackage{amsthm}
\usepackage{amssymb}
\usepackage{color}
\usepackage[colorlinks=true,urlcolor=blue,citecolor=blue,linkcolor=blue]{hyperref}

\DeclareMathOperator{\Tr}{Tr}

\newcommand{\ket}[1]{\left|#1\right\rangle}
\newcommand{\bra}[1]{\left\langle#1\right|}
\DeclareMathOperator{\D}{d}

\DeclareMathOperator{\sech}{sech}

\newtheorem{statement}{Statement}
\newtheorem{definition}{Definition}

\begin{document}
	\title{Latent optical nonclassicality of conditionally prepared states}
	
	\author{V. S. Kovtoniuk}
	\affiliation{Quantum Optics and Quantum Information Group, Bogolyubov Institute for Theoretical Physics of the National Academy of Sciences of Ukraine, Vulytsia Metrolohichna 14b, 03143 Kyiv, Ukraine}
	\author{A. B. Klimov}
	\affiliation{Departamento de F\'isica, Universidad de Guadalajara, 44420 Guadalajara, Jalisco, Mexico}%
	\author{A. A. Semenov}
	\affiliation{Quantum Optics and Quantum Information Group, Bogolyubov Institute for Theoretical Physics of the National Academy of Sciences of Ukraine, Vulytsia Metrolohichna 14b, 03143 Kyiv, Ukraine}
	\affiliation{Department of Theoretical and Mathematical Physics, Kyiv Academic University, Boulevard Vernadskogo  36, 03142  Kyiv, Ukraine}
	
\begin{abstract}

The lack of information obtained from informationally incomplete quantum measurements can prevent the detection of quantum resources, such as optical nonclassicality. 
We develop a technique that overcomes this limitation for single-mode quantum states conditionally prepared through measurements on another mode of a two-mode state.
This task is performed by testing steering of latent nonclassicality---a class of correlations beyond quantum entanglement and quantum discord---for which we introduce a rigorous description.
\end{abstract}
	\maketitle


\section{Introduction}
\label{Sec:Intro}

Ever since the early days of quantum optics, it has been observed that all measurements performed on statistical mixtures of coherent states can be explained without the need to quantize the electromagnetic field.
This is not the case for many archetypal quantum states of light, such as Fock states, quadrature-squeezed states, quantum superpositions of coherent states, etc.
These states of light, whose Glauber-Sudarshan $P$ functions \cite{glauber63c,sudarshan63} cannot be interpreted as probability distributions, are considered in quantum optics as nonclassical  \cite{titulaer65,mandel86,mandel_book,vogel_book,agarwal_book,Schnabel2017,sperling2018a,sperling2018b,sperling2020}.
Optical nonclassicality encompasses a range of phenomena, and various techniques are used to test and quantify it; see, e.g., Refs.~\cite{Stoler1970,Stoler1971,mandel79,reid1986,Wu1986,Wu1987,Hillery1987,Lee1991,agarwal92,agarwal93,klyshko1996,vogel00,richter02,Asboth2005,kiesel08,rivas2009,kiesel10,kiesel11b,sperling12c,bartley13,sperling13b,Agudelo2013,luis15,Miranowicz2015a,Miranowicz2015b,park2015a,park2015b,Yadin2019,Luo2019,bohmann2020,bohmann2020b,Perina2020,semenov2021,Innocenti2023,Fiurasek2024,Kovtoniuk2024}.
Notably, the definitions of optical nonclassicality and nonclassicality in quantum information are maximally inequivalent \cite{ferraro12}. 

Optical nonclassicality, as an example of a broader class of phenomena related to quantum coherence \cite{Streltsov2017,sperling2018a}, is often discussed in a more general context of the quantum resource theories \cite{Horodecki2013,Chitambar2019}. 
Specifically, nonclassical states are the resource for generating quantum entanglement using only passive linear optics \cite{Asboth2005,Miranowicz2015b}. 
They are also an indispensable resource for the computational complexity in several quantum-computing models using bosonic fields---e.g., optical networks \cite{rahimi-keshari16} and quantum kernel methods \cite{Chabaud2024}. 

The preparation of nonclassical states is an essential part of quantum-optical experiments.
In this context, protocols of conditional preparation (see, e.g., Refs.~\cite{hong86,Watanabe1988,DAriano1999,Foster2000,Lvovsky2001,Fiurasek2001,Laurant2003,Laurat2004,Zavatta2004,Zavatta2004a,ourjoumtsev06,zavatta07,kiesel08,Branczyk2010,Agudelo2017,Walschaers2020,Davis2021}) play a crucial role.
Let us consider two parties, Alice and Bob, sharing a two-mode state described by a density operator $\hat{\rho}$.
Alice randomly chooses a device setting related to an observable to be measured and then performs a measurement on her mode to prepare a quantum state of Bob's mode.
This measurement is described by a positive operator-valued measure (POVM) $\hat{\Pi}_{\mathrm{A}}(A|a)$, where $A$ is the measurement outcome and $a$ is a discrete parameter controlling the device setting.
The density operator of Bob's mode conditioned by the outcome $A$ with the setting $a$ reads
	\begin{align}		
        \hat{\rho}_{\mathrm{B}}(A,a)=\mathcal{N}\Tr_{\mathrm{A}}\left[\hat{\rho}\,\hat{\Pi}_{\mathrm{A}}(A|a)\right],
	\end{align}
where $\Tr_{\mathrm{A}}$ denotes the partial tracing over Alice's mode and $\mathcal{N}$ is the normalization factor.

Bob's task is to determine whether at least one of the obtained states $\hat{\rho}_{\mathrm{B}}(A,a)$ is nonclassical, i.e., cannot be represented as a statistical mixture of coherent states $\ket{\alpha}$, where $\alpha\in\mathbb{C}$ is the coherent amplitude.
To do this, he performs the measurement described by the POVM $\hat{\Pi}_{\mathrm{B}}(B|b)$ with $B$ and $b$ being the measurement outcome and the device setting, respectively.
Let us assume that this measurement is not informationally-complete. 
This implies that the conditional probability distributions $\mathcal{P}_{\mathrm{B}}(B|b;A,a)=\Tr_{\mathrm{B}}\left[\hat{\rho}_{\mathrm{B}}(A,a)\hat{\Pi}_{\mathrm{B}}(B|b)\right]$ do not suffice to uniquely determine the density operators $\hat{\rho}_{\mathrm{B}}(A,a)$.

In particular, informational incompleteness may prevent the detection of nonclassicality.
In such cases, which we refer to as latent optical nonclassicality, measurement statistics obtained from a nonclassical state $\hat{\rho}_{\mathrm{B}}(A,a)$ can still be reproduced with a statistical mixture of coherent states.
This means that there exists such $\varrho_{\mathrm{B}}(\alpha|A,a)\geq0$ that $\mathcal{P}_{\mathrm{B}}(B|b;A,a)$
can be represented as a convex combination,
\begin{align}
	\label{Eq:NoSteeringCondition1}
	\mathcal{P}_{\mathrm{B}}(B|b;A,a) = \int_\mathbb{C} d^2 \alpha \varrho_{\mathrm{B}}(\alpha|A,a) \Pi_{\mathrm{B}}(B|b;\alpha),
\end{align}
of coherent-state probability distributions $\Pi_{\mathrm{B}}(B|b;\alpha)=\bra{\alpha}\hat{\Pi}_{\mathrm{B}}(B|b)\ket{\alpha}$, see Refs.~\cite{semenov2021,Innocenti2022,Innocenti2023,Kovtoniuk2024,Fiurasek2024}.
In particular, Eq.~(\ref{Eq:NoSteeringCondition1}) is fulfilled if $\varrho_{\mathrm{B}}(\alpha|A,a)$ is the $P$ function of the state $\hat{\rho}_{\mathrm{B}}(A,a)$.
However, for nonclassical states, the $P$ functions do  not satisfy the restriction of being positive-semidefinite distributions.

In this paper, we describe a technique that can reveal latent optical nonclassicality of Bob's states $\hat{\rho}_{\mathrm{B}}(A,a)$ even if $\varrho_{\mathrm{B}}(\alpha|A,a)\geq0$ in Eq.~(\ref{Eq:NoSteeringCondition1}) exists. 
Our method is based on testing a particular class of correlations for the two-mode states $\hat{\rho}$---steering of latent nonclassicality (SLN)---for which we provide a rigorous description.
These correlations, similar to those considered in Refs.~\cite{Agudelo2013,Koehnke2021,Kovtoniuk2022}, exist for quantum states with no entanglement and zero quantum discord.

The rest of the paper is organized as follows.
In Sec.~\ref{Sec:LHCS-model} we derive the local hidden classical states (LHCS) model based on the assumption that the measurement on Alice's side does not lead to nonclassicality on Bob's side.
In Sec.~\ref{Sec:NonclSteer} we introduce the notion of SLN and the related notion of steering of explicit nonclassicality (SEN).
Inequalities for detecting violations of the LHCS model and, more specifically, SLN are introduced in Sec.~\ref{Sec:Inequal}. 
In Sec.~\ref{Sec:Example} we discuss an example demonstrating SLN for a separable state with zero quantum discord.
Summary and concluding remarks are given in Sec.~\ref{Sec:Concl}.


\section{Local hidden classical states model}
\label{Sec:LHCS-model}

Let us suppose that every conditional state $\hat{\rho}_\mathrm{B}(A,a)$ on Bob's side is classical.
This implies that Eq.~(\ref{Eq:NoSteeringCondition1}), considered as an integral equation for the unknown function $\varrho_{\mathrm{B}}(\alpha|A,a)$, has nonnegative solutions.
Non-negative solutions, corresponding to a particular setting $a$, can be combined to construct unconditional probability distribution
    \begin{align}\label{Eq:UnconditionalPD}
        \varrho_\mathrm{B}(\alpha)=\sum_{A} \varrho_\mathrm{B}(\alpha|A,a) \mathcal{P}_{\mathrm{A}}(A|a),
    \end{align}
where $\mathcal{P}_{\mathrm{A}}(A|a) = \Tr \left[ \hat{\rho} \hat{\Pi}_{\mathrm{A}}(A|a) \otimes \hat{I}_{\mathrm{B}} \right]$ is the probability distribution of $A$ given the device setting $a$.
If the states $\hat{\rho}_\mathrm{B}(A,a)$ are classical, their $P$ functions, $P_\mathrm{B}(\alpha|A,a)$, should be among the solutions to Eq.~(\ref{Eq:NoSteeringCondition1}).
Substituting this particular solution into Eq.~(\ref{Eq:UnconditionalPD}), we conclude that in this case $\varrho_\mathrm{B}(\alpha)$ is the $P$ function, $P_\mathrm{B}(\alpha)$, of the unconditionally reduced state $\hat{\rho}_{\mathrm{B}}=\Tr_{\mathrm{A}}\hat{\rho}$.
Since $P_\mathrm{B}(\alpha)$ does not depend on the device settings $a$, the functions $\varrho_{\mathrm{B}}(\alpha|A,a)=P_\mathrm{B}(\alpha|A,a)$ should satisfy the condition   
    \begin{align}
        \label{Eq:NoSteeringCondition2}
        \sum_{A} \varrho_\mathrm{B}(\alpha|A,a_i) \mathcal{P}_{\mathrm{A}}(A|a_i) = \sum_{A} \varrho_\mathrm{B}(\alpha|A,a_j) \mathcal{P}_{\mathrm{A}}(A|a_j),
    \end{align}
for all $i$ and $j$, where $a_i$ and $a_j$ are different device settings on Alice's side.
If no function among $\varrho_{\mathrm{B}}(\alpha|A,a)\geq0$ satisfies condition (\ref{Eq:NoSteeringCondition2}), then at least one state $\hat{\rho}_\mathrm{B}(A,a)$ is nonclassical, even if the probability distributions $\mathcal{P}_{\mathrm{B}}(B|b;A,a)$ can be reproduced with statistical mixtures of coherent states.

Let us now multiply both sides of Eq.~(\ref{Eq:NoSteeringCondition1}) by $\mathcal{P}_{\mathrm{A}}(A|a)$.
According to the Bayesian rule, the left-hand side becomes $\mathcal{P}_{\mathrm{A}}(A|a)\mathcal{P}_{\mathrm{B}}(B|b;A,a)=\mathcal{P}(A,B|a,b)$, which is the joint probability distribution of $A$ and $B$ given the device settings $a$ and $b$ on Alice's and Bob's side, respectively.
Introducing the function
    \begin{align}\label{Eq:CHSConditionProof}
        F_{\mathrm{A}}(A|a;\alpha) = \frac{\mathcal{P}_{\mathrm{A}}(A|a) \varrho_\mathrm{B}(\alpha|A;a)}{\varrho_\mathrm{B}(\alpha)}\geq 0,
    \end{align}
satisfying the normalization condition $\sum_{A}F_{\mathrm{A}}(A|a;\alpha)=1$, we arrive at the expression
    \begin{align}\label{Eq:NonclassicalSteering}
        \mathcal{P}(A,B|a,b) = \int_\mathbb{C} d^2 \alpha\, \varrho_\mathrm{B}(\alpha) F_{\mathrm{A}}(A|a;\alpha) \Pi_{\mathrm{B}}(B|b;\alpha),
    \end{align}
which establishes the main model of our study.
Equation~(\ref{Eq:NonclassicalSteering}) can be fulfilled only if (1) there exists $\varrho_\mathrm{B}(\alpha|A,a)\geq0$ and (2) $\varrho_\mathrm{B}(\alpha)$ does not depend on $a$, i.e., condition (\ref{Eq:NoSteeringCondition2}) is satisfied.

\begin{figure}[ht!!]
\includegraphics[width=\linewidth]{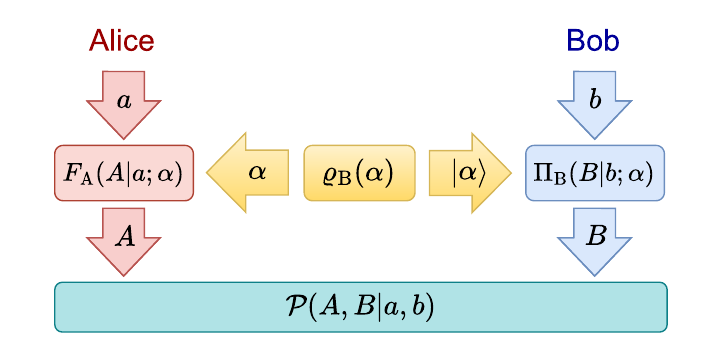}
\caption{\label{Fig:LHCS} The LHCS model. The state is assumed to be characby the random coherent amplitude $\alpha$ distributed according to $\varrho_\mathrm{B}(\alpha)$.
Bob obtains the coherent state $\ket{\alpha}$ and makes with it the measurement described by the coherent-state probability distribution  $\Pi_{\mathrm{B}}(B|b;\alpha)$.
His outcome is $B$ given the device setting $b$.
Alice gets the value of the coherent amplitude $\alpha$ and generates the outcome $A$ given the device setting $a$ according to the probability distribution (the device response function) $F_{\mathrm{A}}(A|a;\alpha)$.
The resulting probability distribution $\mathcal{P}(A,B|a,b)$ is given by Eq.~(\ref{Eq:NonclassicalSteering}).}
\end{figure}

Considering $F_{\mathrm{A}}(A|a;\alpha)$ as a device response function, we can now interpret Eq.~(\ref{Eq:NonclassicalSteering}) in the framework of the LHCS model, which can be described as follows; cf. Fig.~\ref{Fig:LHCS}.
A two-mode state is supposed to be characterized by a random value of the hidden parameter, which is the coherent amplitude $\alpha$, distributed according to $\varrho_{\mathrm{B}}(\alpha)$.
The random value of $A$ on Alice's side obeys the conditional probability distribution described by the response function $F_{\mathrm{A}}(A|a;\alpha)$. 
Bob obtains the coherent state $\ket{\alpha}$ and performs the measurement with it, obtaining the value of $B$ given the device setting $b$.
If for the given $\mathcal{P}(A,B|a,b)$ the LHCS model fails, i.e., there exist no functions $\varrho_\mathrm{B}(\alpha)\geq0$ and $F_{\mathrm{A}}(A|a;\alpha)\geq0$ such that Eq.~(\ref{Eq:NonclassicalSteering}) is fulfilled, then at least one conditional state on Bob's side is nonclassical even if there exists $\varrho_\mathrm{B}(\alpha|A,a)\geq0$.

\section{Steering of explicit and latent nonclassicality}
\label{Sec:NonclSteer}

As concluded in the previous section, a violation of the LHCS model implies that at least one conditional state on Bob's side is nonclassical.
However, this includes trivial cases.
Let us consider, for example, a scenario where the probability distribution is factorized, $\mathcal{P}(A,B|a,b)=\mathcal{P}_{\mathrm{A}}(A|a)\mathcal{P}_{\mathrm{B}}(B|b)$, and $\mathcal{P}_{\mathrm{B}}(B|b)$ cannot be reproduced with statistical mixtures of coherent states.
In this case, the violation of the LHCS model does not give any new information about nonclassicality on Bob's side, nor does it indicate any correlations.

Another example is related to the case when conditional probability distributions on Bob's side, $\mathcal{P}_{\mathrm{B}}(B|b;A,a)$, cannot be represented in the form of Eq.~(\ref{Eq:NoSteeringCondition1}), i.e., they cannot be reproduced with a statistical mixture of coherent states.
Again, the violation of the LHCS model, which is necessary in this case, does not give any new information about nonclassicality on Bob's side.
For such scenarios, we introduce the concept of  \textit{steering of explicit nonclassicality} (SEN).
It is defined as follows.

\begin{definition}
		A given probability distribution $\mathcal{P}(A,B|a,b)$ manifests SEN if two conditions are simultaneously fulfilled: 
		\begin{enumerate}
			\item At least one conditional probability distribution on Bob's side, $\mathcal{P}_{\mathrm{B}}(B|b;A,a)$, cannot be reproduced by a statistical mixture of coherent states, i.e., cannot be represented in the form of Eq.~(\ref{Eq:NoSteeringCondition1});
			
			\item The unconditionally reduced probability distribution on Bob's side, $\mathcal{P}_{\mathrm{B}}(B|b)=\Tr\left[\hat{\rho}_{\mathrm{B}}\hat{\Pi}_{\mathrm{B}}(B|b)\right]$, can be reproduced by a statistical mixture of coherent states.
		\end{enumerate}
\end{definition}

We focus on nontrivial examples of LHCS model violations.
In these cases, additional information about nonclassicality on Bob's side is obtained beyond what is accessible through direct detection.
For such scenarios, we introduce the concept of \textit{steering of latent nonclassicality} (SLN), which we define as follows.

\begin{definition}
	A given probability distribution $\mathcal{P}(A,B|a,b)$ manifests SLN if two conditions are simultaneously fulfilled:
	\begin{enumerate}
		\item $\mathcal{P}(A,B|a,b)$ violates the LHCS model defined by Eq.~(\ref{Eq:NonclassicalSteering});
		
		\item The corresponding conditional probability distributions on Bob's side, $\mathcal{P}_{\mathrm{B}}(B|b;A,a)$, can be reproduced by a statistical mixture of coherent states, i.e., can be represented in the form of Eq.~(\ref{Eq:NoSteeringCondition1}).
	\end{enumerate}	
\end{definition}

The manifestation of SLN indicates that although all conditional probability distributions on Bob's side can be reproduced with statistical mixtures of coherent states, at least one conditional quantum state on Bob's side is nonclassical.
In other words, SLN reveals latent nonclassicality of conditionally prepared states.
This implies that SLN provides additional information about nonclassicality on Bob's side beyond what can be obtained through direct methods of nonclassicality detection, such as given, e.g., in Refs~\cite{semenov2021,Kovtoniuk2024}.

For the probability distributions $\mathcal{P}(A,B|a,b)$ that manifest SEN or SLN the unconditionally reduced probability distribution $\mathcal{P}_{\mathrm{B}}(B|b)$ can be reproduced with a statistical mixture of coherent states.
In these cases, the violation of the LHCS model implies that Alice's measurements create such nonclassicality on Bob's side, which can be detected with the given measurement settings either directly or through testing SLN. 
An especially interesting case is related to the scenarios in which the unconditionally reduced states on Bob's side, $\hat{\rho}_\mathrm{B}$, are a priori classical.
In such scenarios, SEN and SLN imply that Alice's measurements create nonclassical states on Bob's side.

SEN and SLN represent a class of correlations intrinsic to quantum optical systems.
However, they cannot be considered as purely quantum correlations.
The key issue is that genuine quantum correlations should not arise under the local operations and classical communication (LOCC), which is not the case in our scenario.
One of the reasons for this is that our analysis does not extend to the full quantum state, but is instead constrained by the incomplete information available from the given measurement devices.
An alternative definition of nonclassical steering for Gaussian states, measurements, and operations---designed to comply with the LOCC requirement---is presented in Ref.~\cite{frigerio2021}.

SEN and SLN are related to other types of quantum correlations.
First, the local realistic model---whose violation leads to Bell nonlocality \cite{Brunner2014}---can be obtained from the LHCS model replacing $\alpha$ with arbitrary hidden variables $\omega$ and the coherent-state probability distribution $\Pi_{\mathrm{B}}(B|b;\alpha)$ on Bob's side with an arbitrary non-negative normalized response function $F_{\mathrm{B}}(B|b;\omega)$.
Second, the model of classical optical correlations---whose violation leads to so-called nonclassical correlations \cite{Agudelo2013,Koehnke2021,Kovtoniuk2022}---can be obtained from the LHCS model replacing $\alpha$ with two coherent amplitudes, $\alpha_{\mathrm{A}}$ and $\alpha_{\mathrm{B}}$, on Alice's and Bob's side, respectively, and an arbitrary response function $F_{\mathrm{A}}(A|a; \alpha)$ with the coherent-state probability distribution $\Pi_{\mathrm{A}}(A|a;\alpha_{\mathrm{A}})$ on Alice's side.
Third, the model of local hidden states---whose violation leads to regular quantum steering \cite{Wiseman2007,Saunders2010,Cavalcanti2017,Uola2020}---can be obtained from the LHCS model replacing $\alpha$ with arbitrary hidden variables $\omega$ and $\Pi_{\mathrm{B}}(B|b;\alpha)$ with $\mathcal{P}_{\mathrm{B}}(B|b;\omega) = \Tr\left[ \hat{\rho}(\omega) \hat{\Pi}_{\mathrm{B}}(B|b) \right]$, where $\hat{\rho}(\omega)$ are positive-semidefinite unit-trace operators, representing single-mode states.  

The informational incompleteness of measurements imposes inherent constraints that may fundamentally prevent the detection of SLN.
For example, it can be shown that if a given measurement scheme fails to reveal both nonclassicality on Bob's side and regular quantum steering, it likewise cannot reveal SLN or, more generally, violations of the LHCS model.
In particular, this implies that the measurement schemes employing an on-off detector on Bob's side---designed to register only the presence or absence of photons---is unsuitable for detecting SLN.
For details, see Appendix~\ref{Sec:NoViolations}.


\section{Inequality testing the LHCS model}
\label{Sec:Inequal}

The set of all $\mathcal{P}(A,B|a,b)$ satisfying the LHCS model forms a convex set within the space of corresponding probability distributions. 
This is analogous to the sets of local realistic \cite{Brunner2014}, separable \cite{horodecki09}, and other probability distributions $\mathcal{P}(A,B|a,b)$.
Hence, to test the LHCS model, we can use the supporting hyperplane theorem \cite{boyd_book}, i.e., a technique similar to those leading to a generalized form of Bell inequalities and entanglement witness \cite{Brunner2014,semenov2021,Kovtoniuk2022,horodecki96,terhal99,horodecki09}.
We aim to formulate inequalities for $\mathcal{P}(A,B|a,b)$, witnessing violation of the LHCS model (\ref{Eq:NonclassicalSteering}).

It is convenient to recast the LHCS model in an alternative form similar to the Fine formulation \cite{fine82,kaszlikowski2000,abramsky2011} of the local realistic model, whose violation yields Bell nonlocality. 
Let $m_\mathrm{A}$ be the number of device settings available to Alice.
The symbols $\mathcal{A}_i$ denote the observables on Alice's side corresponding to the setting $a_i$, and $\boldsymbol{\mathcal{A}} = \left( \mathcal{A}_1, \ldots, \mathcal{A}_{m_\mathrm{A}} \right)$ is the set of all these observables.
  
    \begin{statement}
        \label{St:FineLikeFormulation}
        An LHCS model defined by Eq.~(\ref{Eq:NonclassicalSteering}) exists for a given probability distribution $\mathcal{P}(A,B|a,b)$ iff there exists a non-negative joint probability distribution of all Alice's observables $\boldsymbol{\mathcal{A}}$ and the coherent amplitude $\alpha$, $\mathcal{W}(\boldsymbol{\mathcal{A}}, \alpha)$, such that 
            \begin{align}
                \label{Eq:FineCHS}
                \mathcal{P}(A,B & |a_i,b_j) \nonumber\\
                & = \sum_{\boldsymbol{\mathcal{A}}} \int_\mathbb{C} d^2 \alpha \mathcal{W}(\boldsymbol{\mathcal{A}},\alpha) \delta_{A,\mathcal{A}_i} \Pi_{\mathrm{B}}(B|b_j;\alpha).
            \end{align}
    \end{statement}
Here the sum is taken over all values of Alice's observables $\boldsymbol{\mathcal{A}}$, see Appendix~\ref{Sec:LHCS} for details.\footnote{Compare with the Fine formulation of the local realistic model, where one deals with the joint probability distribution of all observables, $\mathcal{W}(\boldsymbol{\mathcal{A}},\boldsymbol{\mathcal{B}})$. Here $\boldsymbol{\mathcal{B}}$ is the set of all Bob's observables.}
To prove this statement we note that if the model (\ref{Eq:NonclassicalSteering}) exists, then the joint probability distribution $\mathcal{W}(\boldsymbol{\mathcal{A}},\alpha)$ can be constructed, for instance, as
    \begin{align}
        \mathcal{W}(\boldsymbol{\mathcal{A}}, \alpha) = \varrho_{\mathrm{B}}(\alpha) \prod_{i = 1}^{m_\mathrm{A}} F_{\mathrm{A}}(\mathcal{A}_i|a_i;\alpha) \geq 0.
    \end{align}
On the other hand, if $\mathcal{W}(\boldsymbol{\mathcal{A}}, \alpha) \geq 0$ exists, then the expressions
    \begin{align}
        & \varrho_{\mathrm{B}}(\alpha) = \sum_{\boldsymbol{\mathcal{A}}} \mathcal{W}(\boldsymbol{\mathcal{A}}, \alpha) \geq 0, \\
        & F_{\mathrm{A}}(A|a_i;\alpha) = \frac{ \sum_{\boldsymbol{\mathcal{A}}} \mathcal{W}(\boldsymbol{\mathcal{A}}, \alpha) \delta_{A,\mathcal{A}_i}}{\varrho_{\mathrm{B}}(\alpha)} \geq 0
    \end{align}
determine the probability density $\varrho_{\mathrm{B}}(\alpha)$ and the function $F_{\mathrm{A}}(A|a_i;\alpha)$.

It is also convenient to use a geometric interpretation of Eq.~(\ref{Eq:FineCHS}).
Without loss of generality, we assume that the discrete values of $A$ and $B$ both belong to subsets of integers.
Let $\boldsymbol{\mathcal{P}}$ be the vector with the components $\mathcal{P}(A,B|a_i,b_j)$ enumerated by the lexicographically ordered multi-index $(A,i,B,j)$.
Similarly, we define the vector $\boldsymbol{D}_{\mathrm{A}}(\boldsymbol{\mathcal{A}})$ with the components of the deterministic probability distribution $\delta_{A, \mathcal{A}_{i}}$, cf. Ref.~\cite{Brunner2014}, on Alice's side and the vector $\boldsymbol{\Pi}_{\mathrm{B}}(\alpha)$ with the components of the coherent-state probability distribution $\Pi_{\mathrm{B}}(B|b_j; \alpha)$ on Bob's side enumerated by the lexicographically ordered multi-indices $(A,i)$ and $(B,j)$, respectively.
In such geometric notations, Eq.~(\ref{Eq:FineCHS}) is rewritten as
    \begin{align}\label{Eq:NonclassicalSteering-Geom}
        \boldsymbol{\mathcal{P}}=\sum_{\boldsymbol{\mathcal{A}}} \int_\mathbb{C} d^2 \alpha \mathcal{W}(\boldsymbol{\mathcal{A}},\alpha)\boldsymbol{M}(\boldsymbol{\mathcal{A}}, \alpha),
    \end{align}
where
    \begin{align}\label{Eq:Xi}
        \boldsymbol{M}(\boldsymbol{\mathcal{A}}, \alpha)=\boldsymbol{D}_{\mathrm{A}}(\boldsymbol{\mathcal{A}})\otimes\boldsymbol{\Pi}_{\mathrm{B}}(\alpha).
    \end{align}
This implies that the probability distribution $\boldsymbol{\mathcal{P}}$ obeys an LHCS model iff it belongs to the convex hull of the probability distributions $\boldsymbol{M}(\boldsymbol{\mathcal{A}}, \alpha)$.
Thus, we can use the supporting hyperplane theorem \cite{boyd_book} and the methods of Refs.~\cite{semenov2021,Kovtoniuk2022,Kovtoniuk2024} to formulate the following statement.

    \begin{statement}
        \label{St:Inequalities}
        An LHCS model defined by Eq.~(\ref{Eq:NonclassicalSteering-Geom}) exists for a given probability distribution $\boldsymbol{\mathcal{P}}$ iff for any vector $\boldsymbol{\lambda}$ the inequality
            \begin{align}
                \label{Eq:Inequalities}
                \boldsymbol{\lambda}\cdot\boldsymbol{\mathcal{P}}\leq \sup_{\boldsymbol{\mathcal{A}},\alpha}\boldsymbol{\lambda}\cdot
                \boldsymbol{M}(\boldsymbol{\mathcal{A}}, \alpha)
            \end{align}
        is satisfied.
    \end{statement}

On the right-hand side of this inequality, the supremum is taken over all values of $\alpha\in\mathbb{C}$ and all possible values of $\boldsymbol{\mathcal{A}}$.
The dimension of the vector $\boldsymbol{\mathcal{P}}$ can be significantly reduced by excluding its dependent components using normalization and no-signaling constraints \cite{cirelson80,Popescu1994}, as is done in Bell-inequality analysis \cite{Pironio2005,Brunner2014}; see also Appendix~\ref{Sec:Independ} for details. 
Let us stress that the left-hand side of this inequality can be directly sampled from the measurement data.
The right-hand side can either be evaluated analytically or sampled  experimentally using coherent states on Bob's side.

If there exist such $\boldsymbol{\lambda}$ that inequality (\ref{Eq:Inequalities}) is violated, then at least one of the conditionally prepared states on Bob's side, $\hat{\rho}_{\mathrm{B}}(A,a)$, is nonclassical.
Moreover, if there exist such $\varrho_{\mathrm{B}}(\alpha|A,a)\geq0$ that Eq.~(\ref{Eq:NoSteeringCondition1}) holds, then the probability distribution $\mathcal{P}(A,B|a,b)$ manifests SLN.
The latter condition can also be checked, for example, with inequalities similar to inequality (\ref{Eq:Inequalities}) formulated for testing nonclassicality, see Refs.~\cite{semenov2021,Kovtoniuk2024}.
Thus, the hyperplane supporting theorem provides a useful tool for testing SLN.


\section{An example}
\label{Sec:Example}

We demonstrate the applicability of our method with phase-randomized two-mode squeezed vacuum states (TMSVS),
   	\begin{align}\label{Eq:TMSVS}
    		\hat{\rho} = \sech^2 r \sum_{n = 0}^{+\infty} \tanh^{2n} r \ket{n, n} \bra{n, n},
    \end{align}
where $r \neq 0$ is the squeezing parameter and $\ket{n,n}$ is the two-mode Fock state with $n$ photons at each mode.
The unconditionally reduced state on Bob's side, $\hat{\rho}_{\mathrm{B}}$, is a thermal state, which is classical.
As discussed in Refs.~\cite{Agudelo2013,Koehnke2021}, despite the fact that states (\ref{Eq:TMSVS}) are not entangled and have zero quantum discord, they still exhibit nonclassical correlations.
We show that the states (\ref{Eq:TMSVS}) can reveal nonclassicality of conditionally prepared states on Bob's side remaining latent due to the informational incompleteness of his measurement.

\begin{figure}[ht!]
    \includegraphics[width=0.95 \linewidth]{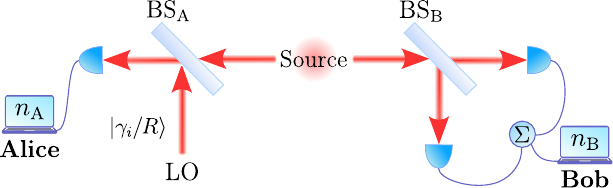}
    \caption{\label{Fig:TMSWS} A measurement scheme for testing SLN of the phase-randomized TMSVS.
    Alice performs displaced on-off detection (also known as unbalanced homodyne detection):
    her mode interferes with the LO in the coherent state $\ket{\gamma_i/R}$ on the beam splitter BS$_{\mathrm{A}}$ having the reflection coefficient $R$ close to zero.
    The obtained signal is analyzed by an on-off detector with the output $n_{\mathrm{A}} \in \left\{ 0, 1 \right\}$.
    The LO amplitude $\gamma_i$ serves as the device setting.
    Bob performs a photocounting measurement with an array of two on-off detectors: the signal field is split by a 50:50 beam splitter BS$_{\mathrm{B}}$ and each output mode is analyzed by an on-off detector.
    The output of Bob's device is the number of triggered detectors, $n_{\mathrm{B}} \in \left\{ 0, 1, 2 \right\}$.}   
\end{figure}

Let us consider the measurement scheme shown in Fig.~\ref{Fig:TMSWS}.    
We deal with a scenario where Alice performs displaced on-off detection (also referred to as unbalanced homodyne detection) \cite{wallentowitz96} with two settings $\gamma_1$ and $\gamma_2$ on her mode.
Here, $\gamma_i$, $i = 1, 2$, is a complex amplitude of a local oscillator (LO) interfering with the signal field through a beam splitter whose transmittance is close to unity.
The measured observable on Alice's side takes two values, $n_\mathrm{A} \in \left\{0, 1\right\}$, which is the output of an on-off detector.
For calculations, we use the POVM
    \begin{align}
        \hat{\Pi}_{\mathrm{A}}(n_{\mathrm{A}}|\gamma_i)=:e^{-(1-n_\mathrm{A})\hat{n}_\mathrm{A}(\gamma_i)}\left(1-e^{-\hat{n}_\mathrm{A}(\gamma_i)}\right)^{n_\mathrm{A}}:.
    \end{align}
Here, $\hat{n}_\mathrm{A}(\gamma_i)=(\hat{a}^\dag_\mathrm{A}-\gamma_i^\ast)(\hat{a}_\mathrm{A}-\gamma_i)$ is the displaced photon-number operator of Alice's mode, $\hat{a}_\mathrm{A}$ and $\hat{a}^\dag_\mathrm{A}$ are the annihilation and creation operators, respectively; and $:\ldots:$ means normal order.
Meanwhile, Bob performs a photocounting measurement with an array of two on-off detectors \cite{paul1996,castelletto2007,schettini2007,blanchet08,achilles03,fitch03,rehacek03} and measures the number of triggered detectors (clicks), $n_\mathrm{B} \in \left\{ 0, 1, 2 \right\}$.
This measurement device has no settings, i.e., technically, its setting has only a single value, which we omit.
The coherent-state probability distributions are given by, cf. Ref.~\cite{sperling12a},
		\begin{align}
			\label{Eq:ArrayPOVM}
			\Pi_{\mathrm{B}}(n_\mathrm{B}|\alpha) = \binom{2}{n_{\mathrm{B}}} e^{-|\alpha|^2(2-n_\mathrm{B})/2} \left(1 - e^{-|\alpha|^2/2}\right)^{n_\mathrm{B}}.
		\end{align}
Detection losses on both sides are included in other losses attributed to the state and described by the efficiencies $\eta_{\mathrm{A}}$ and $\eta_{\mathrm{B}}$.
Technical details are given in Appendix~\ref{Sec:TMSVS}.

\begin{figure}[h]
			\centering
			\includegraphics[width=0.95\linewidth]{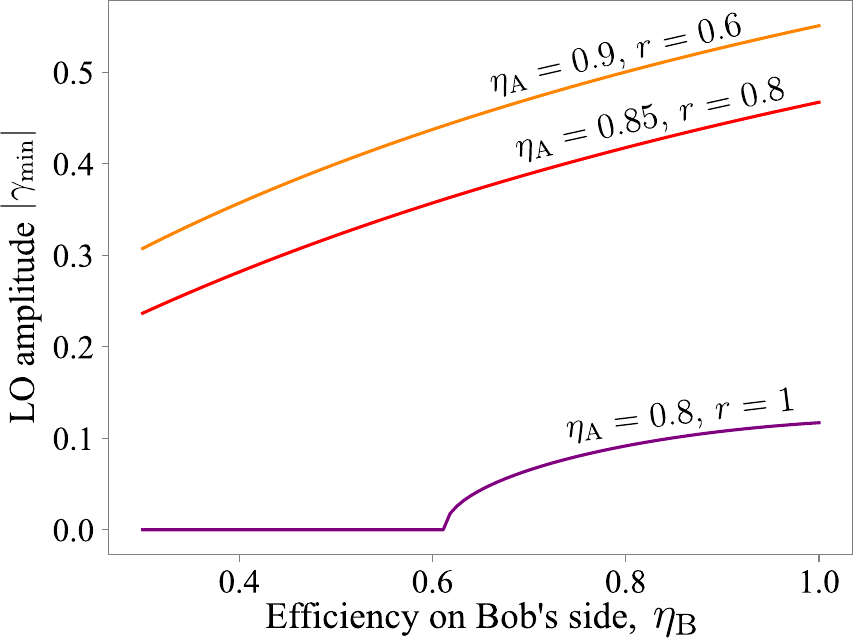}
			\caption{\label{Fig:ConditionalClassicality} Minimal absolute value of the LO amplitude on Alice's side, $|\gamma_{\mathrm{min}}|$, for which the probability distributions on Bob's side, $\mathcal{P}(n_\mathrm{B}|1;\gamma_i)$, can be reproduced with statistical mixtures of coherent states, vs the efficiency $\eta_\mathrm{B}$ for different values of the squeezing parameter is shown for $r$ and the efficiency $\eta_{\mathrm{A}}$. 
            }	
	\end{figure}

Within the described procedure, Alice conditionally prepares four states on Bob's side: $\hat{\rho}_{\mathrm{B}}(n_{\mathrm{A}},\gamma_i)$, where $n_\mathrm{A} \in \left\{0, 1\right\}$, $i=1,2$. 
The informationally incomplete measurement gives him access to partial information about them---the probability distributions $\mathcal{P}_{\mathrm{B}}(n_{\mathrm{B}}|n_{\mathrm{A}},\gamma_i)$.
As shown in Ref.~\cite{Kovtoniuk2024}, they form convex combinations (\ref{Eq:UnconditionalPD}) of the coherent-state probability distributions (\ref{Eq:ArrayPOVM}) iff the inequality
	\begin{align}
			\label{Eq:ArrayClassicality}
			\left[2\mathcal{P}_{\mathrm{B}}(0|n_{\mathrm{A}},\gamma_i) + \mathcal{P}_{\mathrm{B}}(1|n_{\mathrm{A}},\gamma_i)\right]^2 \leq 4\mathcal{P}_{\mathrm{B}}(0|n_{\mathrm{A}},\gamma_i)
	\end{align}
is satisfied.
The probability distributions $\mathcal{P}_{\mathrm{B}}(n_{\mathrm{B}}|0,\gamma_i)$ never violate these inequalities, cf. Appendix~\ref{Sec:TMSVS}.
The probability distributions $\mathcal{P}_{\mathrm{B}}(n_{\mathrm{B}}|1,\gamma_i)$ do not violate them if $|\gamma_i|$ is lower bounded by a value depending on the efficiencies, $|\gamma_i|\geq|\gamma_{\mathrm{min}}(\eta_{\mathrm{A}},\eta_{\mathrm{B}})|$, see Fig.~\ref{Fig:ConditionalClassicality}.
Therefore, in such cases, nonclassicality of Bob's states $\hat{\rho}_{\mathrm{B}}(n_{\mathrm{A}},\gamma_i)$ cannot be detected directly.

Let us show that despite the fact that all probability distributions $\mathcal{P}_{\mathrm{B}}(n_{\mathrm{B}}|n_{\mathrm{A}},\gamma_i)$ on Bob's side can be reproduced with classical electromagnetic fields, the probability distribution $\mathcal{P}(n_{\mathrm{A}},n_{\mathrm{B}}|\gamma_i)$ manifests SLN.
First, we use normalization and no-signaling constraints to reduce the dimension of the vector $\boldsymbol{\mathcal{P}}$ from 12 to eight with the following independent components: $\mathcal{P}(0, 0|\gamma_1)$, $\mathcal{P}(0, 1|\gamma_1)$, $\mathcal{P}(1, 0|\gamma_1)$, $\mathcal{P}(1, 1|\gamma_1)$, $\mathcal{P}(0, 0|\gamma_2)$, $\mathcal{P}(0, 1|\gamma_2)$, $\mathcal{P}_\mathrm{A}(0|\gamma_1)$, and $\mathcal{P}_\mathrm{A}(0|\gamma_2)$; see Appendix~\ref{Sec:Independ}.
Here $\mathcal{P}_\mathrm{A}(0|\gamma_i)=\mathcal{P}(0, 0|\gamma_i)+\mathcal{P}(0, 1|\gamma_i)$ are the marginal probability distributions of Alice's observable, which we use as components of the vector $\boldsymbol{\mathcal{P}}$.
We consider the same components for the vectors $\boldsymbol{M}(\boldsymbol{\mathcal{A}}, \alpha)$.
Next, we note that the coherent-state probability distribution $\Pi_{\mathrm{B}}(n_{\mathrm{B}}|\alpha)$, which depends only on $|\alpha|$, cf. Eq.~(\ref{Eq:ArrayPOVM}), can be reparameterized with $t=e^{-|\alpha|^2/ 2}\in[0,1]$.
The set $\mathcal{C}$ of all vectors $\boldsymbol{M}(\boldsymbol{\mathcal{A}},\alpha)\equiv\boldsymbol{M}(\boldsymbol{\mathcal{A}},t)$ is composed of four open curves in an eight-dimensional space.

We perform the following procedure to find the vector $\boldsymbol{\lambda}$ leading to violation of inequality (\ref{Eq:Inequalities}).
First, we discretize the variable $t$ by $t_k =k/(M - 1)$, where $M=30$ and $k =0 \ldots M-1$ are integers.
Using the \textsf{QuickHull} algorithm \cite{barber96}, we construct the convex hull $\mathcal{H}_M$ of $\boldsymbol{M}(\boldsymbol{\mathcal{A}},t_k)$ for all $\boldsymbol{\mathcal{A}}$ and $t_k$.
This gives us the set of 575,790 vectors $\boldsymbol{\lambda}$ orthogonal to the supporting hyperplanes of $\mathcal{H}_M$.
We select the vector $\boldsymbol{\lambda}$ for which the analog of inequality (\ref{Eq:Inequalities}) for the discrete set of $t_k$ is maximally violated, and then check directly whether the inequality with the same $\boldsymbol{\lambda}$ is violated for the continuous $t$.
Providing an optimization procedure with respect to the LO amplitudes $\gamma_i>|\gamma_{\mathrm{min}}|$, we found examples for which violation of inequality (\ref{Eq:Inequalities}) is statistically significant for the number of sample events $N\gtrsim10^7$; see Table~\ref{Tab:Violation}.
The corresponding \textsf{PYTHON~3} code can be found in the Supplemental Material \cite{supplement}, and technical details are discussed in Appendices~\ref{Sec:Right-Hand-Side} and \ref{Sec:StError}.
Therefore, although the probability distributions $\mathcal{P}(n_{\mathrm{B}}|n_{\mathrm{A}},\gamma_i)$ on Bob's side can be reproduced with classical light, our method indicates that at least one conditionally prepared state $\hat{\rho}_{\mathrm{B}}(n_{\mathrm{A}},\gamma_i)$ is nonclassical.

\begin{table}[h!]
    \centering
        \caption{\label{Tab:Violation} Relative violation (difference of the left- and right-hand sides divided by the standard deviation of the left-hand side) of inequality (\ref{Eq:Inequalities}) for states (\ref{Eq:TMSVS}) for different values of the squeezing parameter $r$, the efficiencies $\eta_{\mathrm{A}}$ and $\eta_{\mathrm{B}}$, and the LO amplitudes $\gamma_i$.
        Here, $N$ is the number of sample events.
        See Appendix~\ref{Sec:StError} for details.}
    \addtolength{\tabcolsep}{3.6pt}
    \begin{tabular}{ccccccc}
    	\hline
    	\hline
    	Case&$r$&$\eta_{\mathrm{A}}$ & $\eta_{\mathrm{B}}$ & $\gamma_1$ & $\gamma_2$ & Relative violation\\
    	\hline
    	(A)&0.6 & 0.9 & 0.75 & 0.54 & 1.04 & $2.6 \times10^{-3} \sqrt{N}$ \\
    	\hline
    	(B)&0.8 & 0.85 & 0.5 & 0.44 & 1.04 & $4.1 \times 10^{-3} \sqrt{N}$ \\
    	\hline
    	(C)&1 & 0.8 & 0.3 & 0.1 & 0.8 & $1.2 \times 10^{-2} \sqrt{N}$ \\
    	\hline
    	\hline
    \end{tabular}
\end{table}


\section{Summary and conclusions}
\label{Sec:Concl}

We have proposed a rigorous description for two classes of correlations of two-mode light beyond quantum entanglement and quantum discord---steering of latent nonclassicality (SLN) and steering of explicit nonclassicality (SEN).
The presence of these correlations indicates that states on Bob's side, conditioned by measurements on Alice's side, are nonclassical, while the unconditionally reduced probability distribution on his side can be reproduced by statistical mixtures of coherent states.
This nonclassicality may not be directly detectable due to the informational incompleteness of Bob's measurements.
This means that although the conditional states are nonclassical, the corresponding probability distributions for the measured observables can be reproduced with classical light.
We have shown that testing for SLN may reveal this latent optical nonclassicality without the need for additional measurement data.

SLN (as well as SEN) implies violations of the LHCS model.
This model is similar to the hidden states model used in the context of standard quantum steering.
In our case, however, the hidden states can only be the coherent states of the radiation field.
Since the probability distributions satisfying the LHCS model form a convex set, SLN can be detected with the technique similar to a generalized form of Bell inequalites or entanglement witness.
This method was demonstrated in an example of a state that is not entangled and whose quantum discord is zero.

V.S.K. and A.A.S. acknowledge supporting this work by the National Research Foundation of Ukraine through the Project Nr. 2020.02/0111. 
A.B.K. acknowledges support from  the Grant CBF2023-2024-50 of CONAHCyT (Mexico).


\appendix    

\section{Conditions preventing LHCS model violations}
\label{Sec:NoViolations}

In this section we prove the fact mentioned at the end of Sec.~\ref{Sec:NonclSteer} about the impossibility of detecting the LHCS model violations with the measurement schemes that fail to detect both nonclassicality on Bob's side and regular quantum steering.
The first condition implies that  for any quantum state one can find $\varrho(\alpha) \geq 0$ such that for all $B$ and $b$
\begin{align}
\label{Eq:Classicality}
\mathcal{P}_{\mathrm{B}}(B|b) = \int_\mathbb{C} \D^2 \alpha \varrho(\alpha) \Pi_{\mathrm{B}}(B|b;\alpha),
\end{align}
cf. Refs.~\cite{semenov2021,Kovtoniuk2024}.
The second condition means that there exist such $\varrho^\prime(\omega) \geq 0$ and $F^\prime(A|a;\omega) \geq 0$ that
\begin{align}
\label{Eq:Steering}
\mathcal{P}(A,B|a,b) = \int_\Omega \D \omega \varrho^\prime(\omega)  F_{\mathrm{A}}^\prime(A|a;\omega) \mathcal{P}_{\mathrm{B}}(B|b;\omega),
\end{align}
where $\mathcal{P}_{\mathrm{B}}(B|b;\omega) = \Tr\left[ \hat{\rho}(\omega) \hat{\Pi}_{\mathrm{B}}(B|b) \right]$ and $\hat{\rho}(\omega)$ are positive-semidefinite unit-trace operators, representing single-mode states; cf. Refs.~\cite{Wiseman2007,Saunders2010,Cavalcanti2017,Uola2020}.

Condition~(\ref{Eq:Classicality}) yields that there exist such $\varrho(\alpha|\omega)\geq 0$ that 
\begin{align}
\mathcal{P}_{\mathrm{B}}(B|b;\omega) =  \int_\mathbb{C} \D^2 \alpha \varrho_{\mathrm{B}}(\alpha|\omega) \Pi_{\mathrm{B}}(B|b;\alpha).
\end{align}
Substituting this expression into Eq.~(\ref{Eq:Steering}) and setting
\begin{align}
& \varrho_{\mathrm{B}}(\alpha) = \int_\Omega \D^2 \omega \varrho^\prime(\omega) \varrho_{\mathrm{B}}(\alpha|\omega), \\
& F_{\mathrm{A}}(A|a;\alpha) = \int_\Omega \D \omega \frac{\varrho^\prime(\omega) \varrho_{\mathrm{B}}(\alpha|\omega)}{\varrho_{\mathrm{B}}(\alpha)} F_{\mathrm{A}}^\prime(A|a;\omega),
\end{align}
we get Eq.~(\ref{Eq:NonclassicalSteering}), implying that the LHCS model holds.

In particular, this means that the LHCS model violations cannot be detected if the measurement on Bob's side is performed using a single on-off detector described by
\begin{align}
\Pi_{\mathrm{B}}(n_{\mathrm{B}}|\alpha)=e^{-(1-n_\mathrm{B})|\alpha|^2}\left(1-e^{-|\alpha|^2}\right)^{n_\mathrm{B}},
\end{align}
where $n_\mathrm{B}=\{0,1\}$.
To prove this, we mention two facts.
First, any probability distribution $\mathcal{P}_{\mathrm{B}}(n_{\mathrm{B}})$ on Bob's side can be simulated by a single coherent state with an amplitude defined as $|\alpha|^2=-\ln \mathcal{P}_{\mathrm{B}}(n_{\mathrm{B}}=0)$.
This means that nonclassicality on Bob's side cannot be identified for any state. 
Second, the detector on Bob's side has only two outcomes and a single setting. 
As demonstrated in Ref.~\cite{Saunders2012}, regular quantum steering cannot be identified in such a scenario either.

\section{Local hidden classical states model}
\label{Sec:LHCS}

In this section, we discuss the formulation of the LHCS model given by Eq.~(\ref{Eq:FineCHS}), i.e., in a way similar to the Fine formulation of the local realistic model \cite{fine82,kaszlikowski2000,abramsky2011}.
First, we note that the joint probability distribution of all Alice's observables $\boldsymbol{\mathcal{A}}$ and the coherent amplitude $\alpha$, $\mathcal{W}(\boldsymbol{\mathcal{A}}, \alpha)$ can be rewritten as $\mathcal{W}(\mathcal{A}_1,\ldots,\mathcal{A}_{m_\mathrm{A}},\alpha)$. 
In such notations, Eq.~(\ref{Eq:FineCHS}) is rewritten as
\begin{widetext}
	\begin{align}
	\mathcal{P}(A,B |a_i,b_j) 
	= \sum_{{\mathcal{A}_1, \ldots, \mathcal{A}_{m_\mathrm{A}}}} \int_\mathbb{C} d^2 \alpha \mathcal{W}(\mathcal{A}_1,\ldots,\mathcal{A}_{m_\mathrm{A}},\alpha) \delta_{A,\mathcal{A}_i} \Pi_{\mathrm{B}}(B|b_j;\alpha).
	\end{align}  
	The sum by $\mathcal{A}_i$ can be calculated explicitly, 
	\begin{align}
	\mathcal{P}(A,B |a_i,b_j) 
	= \int_\mathbb{C} d^2 \alpha
	\underbrace{%
		\sum_{\substack{\mathcal{A}_1, \ldots, \mathcal{A}_{i - 1} \\ \mathcal{A}_{i + 1}, \ldots, \mathcal{A}_{m_{\mathrm{A}}}}}  \mathcal{W}(\mathcal{A}_1,\ldots,\mathcal{A}_{i-1},A,\mathcal{A}_{i+1},\ldots,\mathcal{A}_{m_\mathrm{A}},\alpha)%
	}_{\displaystyle w_i(A,\alpha)}
	\Pi_{\mathrm{B}}(B|b_j;\alpha),
	\end{align} 
	where $w_i(A,\alpha)$ is interpreted as the joint probability distribution of $A$ and $\alpha$ given the device setting $a_i$. 
\end{widetext}

To link this formulation of the LHCS model to the Fine formulation of the local realistic model, we first note that in the latter case we deal with the joint probability distribution of all observables (JPDAO), $\mathcal{W}(\boldsymbol{\mathcal{A}},\boldsymbol{\mathcal{B}})\equiv\mathcal{W}(\mathcal{A}_1,\ldots,\mathcal{A}_{m_\mathrm{A}},\mathcal{B}_1,\ldots,\mathcal{B}_{m_\mathrm{B}})$. 
According to the Fine theorem, if the probability distributions $\mathcal{P}(A,B|a_i,b_j)$ satisfy the local realistic model, they should be marginals of $\mathcal{W}(\boldsymbol{\mathcal{A}},\boldsymbol{\mathcal{B}})$.
This implies
\begin{widetext}
	\begin{align}
	\mathcal{P}(A,B|a_i,b_j) = \sum_{\substack{\mathcal{A}_1, \ldots, \mathcal{A}_{i - 1} \\ \mathcal{A}_{i + 1}, \ldots, \mathcal{A}_{m_{\mathrm{A}}}}} \sum_{\substack{\mathcal{B}_1, \ldots, \mathcal{B}_{j - 1} \\ \mathcal{B}_{j + 1}, \ldots, \mathcal{B}_{m_{\mathrm{B}}}}} \mathcal{W}(\mathcal{A}_1, \ldots, \mathcal{A}_{i-1}, A, \mathcal{A}_{i+1}, \ldots, \mathcal{A}_{m_{\mathrm{A}}}, \mathcal{B}_1, \ldots, \mathcal{B}_{j-1}, B, \mathcal{B}_{j+1}, \ldots, \mathcal{B}_{m_{\mathrm{B}}})
	\end{align}
	or, alternatively,
	\begin{align}
	\mathcal{P}(A,B|a_i,b_j)
	= \sum_{\mathcal{A}_1, \ldots, \mathcal{A}_{m_{\mathrm{A}}}} \sum_{\mathcal{B}_1, \ldots, \mathcal{B}_{m_{\mathrm{B}}}} \mathcal{W}(\mathcal{A}_1, \ldots , \mathcal{A}_{m_{\mathrm{A}}}, \mathcal{B}_1, \ldots , \mathcal{B}_{m_{\mathrm{B}}}) \delta_{A, \mathcal{A}_i} \delta_{B, \mathcal{B}_j}
	= \sum_{\boldsymbol{\mathcal{A}}, \boldsymbol{\mathcal{B}}} \mathcal{W}(\boldsymbol{\mathcal{A}}, \boldsymbol{\mathcal{B}}) \delta_{A, \mathcal{A}_i} \delta_{B, \mathcal{B}_j}.
	\end{align}       
\end{widetext}
The LHCS model can be obtained from this expression replacing summation with respect to Bob's observables $\boldsymbol{\mathcal{B}}$ by integration with respect to the coherent amplitude $\alpha$, the JPDAO $\mathcal{W}(\boldsymbol{\mathcal{A}}, \boldsymbol{\mathcal{B}})$ by the joint probability distribution $\mathcal{W}(\boldsymbol{\mathcal{A}}, \alpha)$, and the deterministic probability distribution $\delta_{B, \mathcal{B}_j}$ by the coherent-state probability distribution $\Pi_{\mathrm{B}}(B|b_j;\alpha)$.

\section{No-signaling and normalization conditions}
\label{Sec:Independ}

In this section we show how the dimension of the vector $\boldsymbol{\mathcal{P}}$ with the components $\mathcal{P}(A,B|a,b)$ can be significantly reduced by using no-signaling and normalization constraints, cf. Refs.~\cite{Pironio2005,Brunner2014}.
Let $m_{\mathrm{A}}$ and $m_{\mathrm{B}}$ be the number of settings on Alice's and Bob's side, respectively.
We also take the number of possible values of $A$ and $B$ to be $M_{\mathrm{A}}$ and $M_{\mathrm{B}}$.
The number of all components of the vector $\boldsymbol{\mathcal{P}}$ is
\begin{align}
\dim \boldsymbol{\mathcal{P}}_{\textrm{full}}=m_{\mathrm{A}}m_{\mathrm{B}}M_{\mathrm{A}}M_{\mathrm{B}}.
\end{align}
However, not all of them are independent.

First, the normalization condition,
\begin{align}\label{Eq:Norm}
\sum_{A,B} \mathcal{P}(A,B|a,b) = 1,
\end{align}
imposes $m_{\mathrm{A}}m_{\mathrm{B}}$ constraints for the probability distributions $\mathcal{P}(A,B|a,b)$.
This means that the number of independent components is reduced to $m_{\mathrm{A}}m_{\mathrm{B}}(M_{\mathrm{A}}M_{\mathrm{B}}-1)$.
Second, the marginal probability distributions, 
\begin{align}
\mathcal{P}_{\mathrm{A}}(A|a)=\sum_{B}\mathcal{P}(A,B|a,b), \label{Eq:PA}\\
\mathcal{P}_{\mathrm{B}}(B|b)=\sum_{A}\mathcal{P}(A,B|a,b), \label{Eq:PB}
\end{align}
should not depend on the settings $b$ and $a$, respectively.
This leads to no-signaling conditions for the probability distribution $\mathcal{P}(A,B|a,b)$,
\begin{align}
\sum_{B} \mathcal{P}(A,B|a,b) = \sum_{B} \mathcal{P}(A,B|a, b^\prime),\quad b\neq b^\prime,\label{Eq:NS-1}\\
\sum_{A} \mathcal{P}(A,B|a,b) = \sum_{A} \mathcal{P}(A,B|a^\prime,b),\quad a\neq a^\prime.\label{Eq:NS-2}
\end{align}
Nevertheless, not all of them are independent due to the normalization conditions.

For given $b$ and $b^\prime$ there are $m_{\mathrm{A}}M_{\mathrm{A}}$ constraints (\ref{Eq:NS-1}).
However, due to $m_{\mathrm{A}}$ normalization conditions for $\mathcal{P}_{\mathrm{A}}(A|a)$, 
\begin{align}\label{Eq:Norm-A}
\sum_A\mathcal{P}_{\mathrm{A}}(A|a)=1,
\end{align}
which directly follows from Eq.~(\ref{Eq:Norm}),
this number is reduced to $m_{\mathrm{A}}(M_{\mathrm{A}}{-}1)$ independent constraints.
Since the number of distinct pairs $(b,b^\prime)$ with $b{\neq}b^{\prime}$ is $(m_{\mathrm{B}}{-}1)$, the total number of independent constraints (\ref{Eq:NS-1}) is $m_{\mathrm{A}}(M_{\mathrm{A}}{-}1)(m_{\mathrm{B}}{-}1)$.
Similarly, the number of independent constraints (\ref{Eq:NS-2}) is $m_{\mathrm{B}}(M_{\mathrm{B}}{-}1)(m_{\mathrm{A}}{-}1)$.
This implies that the dimension of the vector $\boldsymbol{\mathcal{P}}$ is reduced to
\begin{align}
\dim &\boldsymbol{\mathcal{P}}=m_{\mathrm{A}}m_{\mathrm{B}}(M_{\mathrm{A}}M_{\mathrm{B}}-1)\\
&-m_{\mathrm{A}}(M_{\mathrm{A}}{-}1)(m_{\mathrm{B}}{-}1)-m_{\mathrm{B}}(M_{\mathrm{B}}{-}1)(m_{\mathrm{A}}{-}1)\nonumber\\
&=m_{\mathrm{A}}m_{\mathrm{B}}(M_{\mathrm{A}}M_{\mathrm{B}}-M_{\mathrm{A}}-M_{\mathrm{B}}+1)\nonumber\\
&+m_{\mathrm{A}}(M_{\mathrm{A}}-1)+m_{\mathrm{B}}(M_{\mathrm{B}}-1).\nonumber
\end{align}
For the analysis one can consider only independent components of the vector $\mathcal{\boldsymbol{P}}$, including different linear combinations of $\mathcal{P}(A,B|a,b)$, e.g. marginal probability distributions (\ref{Eq:PA}) and (\ref{Eq:PB}).

In the example considered in Sec.~\ref{Sec:Example}, we have $M_{\mathrm{A}}{=}2$, $m_{\mathrm{A}}{=}2$, $M_{\mathrm{B}}{=}3$, and $m_{\mathrm{B}}{=}1$.
This implies $\dim\boldsymbol{\mathcal{P}}{=}8$.
We chose the following independent components of the vector $\boldsymbol{\mathcal{P}}$: $\mathcal{P}(0, 0|\gamma_1)$, $\mathcal{P}(0, 1|\gamma_1)$, $\mathcal{P}(1, 0|\gamma_1)$, $\mathcal{P}(1, 1|\gamma_1)$, $\mathcal{P}(0, 0|\gamma_2)$, $\mathcal{P}(0, 1|\gamma_2)$, $\mathcal{P}_\mathrm{A}(0|\gamma_1)$, and $\mathcal{P}_\mathrm{A}(0|\gamma_2)$.
Other probabilities can be recovered from these components as
\begin{align}
\label{Eq:P02}
& \mathcal{P}(0,2|\gamma_i) = \mathcal{P}_\mathrm{A}(0|\gamma_i) {-} \mathcal{P}(0,0|\gamma_i) {-} \mathcal{P}(0,1|\gamma_i), \\
& \mathcal{P}(1,2|\gamma_1) = 1 {-} \mathcal{P}_\mathrm{A}(0|\gamma_1) {-} \mathcal{P}(1,0|\gamma_1) {-} \mathcal{P}(1,1|\gamma_1), \\
& \mathcal{P}(1,n|\gamma_2) = \mathcal{P}(0,n|\gamma_1) {+} \mathcal{P}(1,n|\gamma_1) {-} \mathcal{P}(0,n|\gamma_2).
\end{align}
We used the vector $\boldsymbol{\mathcal{P}}$ with these independent components to test the LHCS model.

\section{Phase-randomized two-mode squeezed vacuum state}
\label{Sec:TMSVS}

In this section, we represent the independent components of $\boldsymbol{\mathcal{P}}$ in the scenario where the source irradiates a phase-randomized two-mode squeezed vacuum state (TMSVS), and Alice and Bob analyze it using displaced on-off detection (also known as unbalanced homodyne detection) and an array of two on-off detectors, respectively.
Here we also prove that the probability distributions $\mathcal{P}_{\mathrm{B}}(n_{\mathrm{B}}|0,\gamma_i)$ of the states with $n_\mathrm{A} = 0$ can be reproduced with statistical mixtures of coherent states.
For the analysis of nonclassicality and SLN, we assume that all losses, including detection losses, are attributed to the quantum state.
However, it is more convenient to provide calculations for the state given by Eq.~(\ref{Eq:TMSVS}) and include the losses (detection efficiencies $\eta_\mathrm{A}$ and $\eta_\mathrm{B}$) in the POVMs.

The POVM element for displaced on-off detection related to the no-count event on Alice's side with the efficiency $\eta_{\mathrm{A}}$ is given by
\begin{align}
\hat{\Pi}_\mathrm{A}&(0|\gamma_i; \eta_\mathrm{A}) \nonumber \\
& = \frac{1}{\pi \left(1 -\eta_\mathrm{A}\right)} \int_\mathbb{C} \D^2 \beta e^{-\frac{|\beta \sqrt{\eta_\mathrm{A}} - \gamma_i|^2}{1 - \eta_\mathrm{A}}} \ket{\beta} \bra{\beta}. 
\end{align}
The POVM elements for the array of two on-off detectors related to the no-count and single-count events on Bob's side with the efficiency $\eta_{\mathrm{B}}$ reads
\begin{align}
& \hat{\Pi}_\mathrm{B}(0|\eta_\mathrm{B}) = \sum_{n = 0}^\infty (1 - \eta_\mathrm{B})^n \ket{n} \bra{n}, \\
& \hat{\Pi}_\mathrm{B}(1|\eta_\mathrm{B}) \nonumber \\
& = 2 \sum_{n = 1}^\infty \left[ \left(1 - \frac{\eta_\mathrm{B}}{2} \right)^n - \left(1 - \eta_\mathrm{B}\right)^n \right] \ket{n} \bra{n}.
\end{align}
Applying Born's rule
\begin{align}
\mathcal{P}(n_{\mathrm{A}},n_{\mathrm{B}}|\gamma_i)=
\Tr\left[\hat{\rho}\,\hat{\Pi}_\mathrm{A}(n_{\mathrm{A}}|\gamma_i; \eta_\mathrm{A})\hat{\Pi}_\mathrm{B}(n_{\mathrm{B}}|\eta_\mathrm{B}) \right],
\end{align}
we arrive at the independent components of $\boldsymbol{\mathcal{P}}$,
\begin{align}
\label{Eq:P00}
& \mathcal{P}(0, 0|\gamma_i) = \sech^2 r \, f(1 - \tanh^2 r (1 - \eta_\mathrm{B});\gamma_i;\eta_\mathrm{A}), \\
\label{Eq:P01}
& \mathcal{P}(0, 1|\gamma_i) = 2  \left[ \sech^2 r f(\sech^2 r + \frac{\eta_\mathrm{B}}{2} \tanh^2 r ;\gamma_i;\eta_\mathrm{A})  \right. \nonumber \\
& \left. - \mathcal{P}(0, 0|\gamma_i) \right], \\
& \mathcal{P}(1, 0|\gamma_i) = \mathcal{P}_\mathrm{B}(0) - \mathcal{P}(0, 0|\gamma_i), \\
& \mathcal{P}(1, 1|\gamma_i) = \mathcal{P}_\mathrm{B}(1) - \mathcal{P}(0, 1|\gamma_i), \\
\label{Eq:P0A}
& \mathcal{P}_\mathrm{A}(0|\gamma_i) = \frac{1}{1 + \eta_\mathrm{A} \sinh^2 r} \exp\left[-\frac{|\gamma_i|^2}{1 + \eta_\mathrm{A} \sinh^2 r}\right].
\end{align}
Here we use the notation
\begin{align}
& f(s;\gamma_i;\eta_\mathrm{A}) = \frac{1}{s + \eta_\mathrm{A} (1 - s)} \exp\left[-\frac{s |\gamma_i|^2}{s + \eta_\mathrm{A}(1 - s)}\right], 
\end{align}
and
\begin{align}
& \mathcal{P}_\mathrm{B}(0) = \frac{\sech^2 r}{1 - \tanh^2 r (1 - \eta_\mathrm{B})}, \\
& \mathcal{P}_\mathrm{B}(1) = \frac{2 \eta_\mathrm{B} \sinh^2 r}{(1 + \eta_\mathrm{B} \sinh^2 r) (2 + \eta_\mathrm{B} \sinh^2 r)}
\end{align}
are the marginal probabilities for the observable values on Bob's side.

Let us now prove that the probability distributions $\mathcal{P}_{\mathrm{B}}(n_{\mathrm{B}}|0,\gamma_i)$ of the states conditioned on Alice's outcome $n_\mathrm{A} = 0$ can be reproduced with statistical mixtures of coherent states.
This is satisfied if inequality (\ref{Eq:ArrayClassicality}) is fulfilled; see Ref.~\cite{Kovtoniuk2024}.
The expressions involving the conditional probabilities in this inequality read
\begin{align}\label{Eq:PB00}
& \mathcal{P}_\mathrm{B}(0|0;\gamma_i) = \frac{1 + \eta_\mathrm{A} \sinh^2 r}{\cosh^2 r} e^{-s_0 |\gamma_i|^2},
\end{align}
\begin{align}
[ 2 \mathcal{P}_\mathrm{B}(0|0;\gamma_i) &+ \mathcal{P}_\mathrm{B}(1|0;\gamma_i) ]^2 \nonumber \\
& = 2 \frac{1 + \eta_\mathrm{A} \sinh^2 r}{\cosh^2 r} e^{-s_1 |\gamma_i|^2},\label{Eq:PB0010}
\end{align}
where
\begin{align}
\label{Eq:s0}
& s_0 = \frac{C}{\sech^2 r + [\eta_\mathrm{B} + \eta_\mathrm{A}(1 - \eta_\mathrm{B})] \tanh^2 r}, \\
\label{Eq:s1}
& s_1 = \frac{C}{\sech^2 r + \frac{1}{2} [\eta_\mathrm{B} + \eta_\mathrm{A}(2 - \eta_\mathrm{B})] \tanh^2 r}, \\
& C = \frac{\eta_\mathrm{A} \eta_\mathrm{B} \sinh^2 r}{1 + \eta_\mathrm{A} \sinh^2 r}.
\end{align}
Substituting Eqs.~(\ref{Eq:PB00}) and (\ref{Eq:PB0010}) in inequality (\ref{Eq:ArrayClassicality}) yields
\begin{align}
\label{Eq:Classicality2}
\frac{1 + \eta_\mathrm{A} \sinh^2 r}{\cosh^2 r} e^{-|\gamma_i|^2 (s_1 - s_0)} \leq 1.
\end{align}
Since the denominator of $s_0$ in Eq. (\ref{Eq:s0}) is not smaller than that of $s_1$ in Eq. (\ref{Eq:s1}), we have $s_1 \geq s_0$, and hence $e^{-|\gamma_i|(s_1 - s_0)} \leq 1$.
The factor before the exponent in Eq.~(\ref{Eq:Classicality2}) is also not greater than 1.
Thus, the statistics of Bob's conditional state when $n_\mathrm{A} = 0$ is always classical.

\section{Right-hand side of inequality testing the LHCS model}
\label{Sec:Right-Hand-Side}

In this section we discuss the computational details of finding the supremum in inequality~(\ref{Eq:Inequalities}) given the vector $\boldsymbol{\lambda}$. 
The right-hand side of this inequality reads
\begin{align}
\label{Eq:RHS}
\sup_{\boldsymbol{\mathcal{A}}, \alpha} \boldsymbol{\lambda} \cdot \boldsymbol{M}(\boldsymbol{\mathcal{A}}, \alpha).
\end{align}
In the setup considered in the main text, the variable $\boldsymbol{\mathcal{A}} = (\mathcal{A}_1, \mathcal{A}_2)$ takes four values, $(0, 0)$, $(0, 1)$, $(1, 0)$ and $(1, 1)$.
We parameterize the POVM symbol of Bob's measurement with $t = e^{-|\alpha|^2 / 2}$ to obtain
\begin{align}
& \Pi(0|\alpha) = \Pi_0(t) = t^2, \\
& \Pi(1|\alpha) = \Pi_1(t) = 2 t (1 - t).
\end{align}
Then the values of $\boldsymbol{M}(\boldsymbol{\mathcal{A}}, \alpha) \equiv \boldsymbol{M}(\mathcal{A}_1, \mathcal{A}_2, t)$ are given by
\begin{align}
& \boldsymbol{M}(0, 0, t) = \left( \Pi_0(t), \Pi_1(t), 0 , 0, \Pi_0(t), \Pi_1(t), 1, 1 \right), \\
& \boldsymbol{M}(0, 1, t) = \left( \Pi_0(t), \Pi_1(t), 0 , 0, 0, 0, 1, 0 \right), \\
& \boldsymbol{M}(1, 0, t) = \left( 0, 0, \Pi_0(t), \Pi_1(t), \Pi_0(t), \Pi_1(t), 0, 1 \right), \\
& \boldsymbol{M}(1, 1, t) = \left( 0, 0, \Pi_0(t), \Pi_1(t), 0, 0, 0, 0 \right).
\end{align}
This allows us to rewrite Eq. (\ref{Eq:RHS}) as
\begin{align}
\label{Eq:RHS2}
& \max_{\mathcal{A}_1, \mathcal{A}_2} \sup_{t \in [0, 1]} \boldsymbol{\lambda} \cdot \boldsymbol{M}(\mathcal{A}_1, \mathcal{A}_2, t).
\end{align}
For each pair of $\mathcal{A}_1$, $\mathcal{A}_2$, the expression $\boldsymbol{\lambda} \cdot \boldsymbol{M}(\mathcal{A}_1, \mathcal{A}_2, t)$ is a quadratic function of $t$.

Let us define the function    
\begin{align}
q(a; b; c) & = \sup_{t \in [0, 1]} a t^ 2 + b t + c,
\end{align}
which can be easily computed analytically as the global maximum of a quadratic polynomial of $t\in[0,1]$.
Then for the given $\mathcal{A}_1$ and $\mathcal{A}_2$, we have   
\begin{align}\label{Sup2}
\sup_{t \in [0, 1]} \boldsymbol{\lambda} \cdot \boldsymbol{M}(\mathcal{A}_1, \mathcal{A}_2, t) 
= q(a_{\mathcal{A}_1, \mathcal{A}_2};b_{\mathcal{A}_1, \mathcal{A}_2};c_{\mathcal{A}_1, \mathcal{A}_2}).
\end{align}
Here, we use the notations
\begin{align}
& a_{00} = \lambda_{00}^1 - 2 \lambda_{01}^1 + \lambda_{00}^2 - 2 \lambda_{01}^2, \\
& b_{00} = 2 \lambda_{01}^1 + 2 \lambda_{01}^2, \\
& c_{00} = \lambda_0^1 + \lambda_0^2, \\
& a_{01} = \lambda_{00}^1 - 2 \lambda_{01}^1, \\
& b_{01} = 2 \lambda_{01}^1, \\
& c_{01} = \lambda_0^1, \\
& a_{10}  = \lambda_{10}^1 - \lambda_{11}^1 + \lambda_{00}^2 - 2 \lambda_{01}^2, \\
& b_{10} = 2 \lambda_{11}^1 + 2 \lambda_{01}^2, \\
& c_{10} = \lambda_0^2,\\
& a_{11} = \lambda_{10}^1 - 2 \lambda_{11}^1,\\
& b_{11} = 2 \lambda_{11}^1,\\
& c_{11} = 0,
\end{align}
and 
\begin{align}
& \lambda(n_\mathrm{A}, n_\mathrm{B}|\gamma_i) = \lambda_{n_\mathrm{A} n_\mathrm{B}}^i, \\
& \lambda_\mathrm{A}(n_\mathrm{A}|\gamma_i) = \lambda_{n_\mathrm{A}}^i.
\end{align}
Finally, the right-hand side (\ref{Eq:RHS}) of inequality (\ref{Eq:Inequalities}) can be found as the maxima of four numbers given by Eq.~(\ref{Sup2}) for all values of $\boldsymbol{\mathcal{A}} = (\mathcal{A}_1, \mathcal{A}_2)$.

\section{Statistical error and optimization procedure}
\label{Sec:StError}

In this section, we discuss the statistical error that appears when estimating the left-hand side of inequality (\ref{Eq:Inequalities}) based on the measurement data and the procedure optimizing the local-oscillator (LO) amplitudes. 
For brevity, we adopt the same notation for elements of $\boldsymbol{\mathcal{P}}$ that we used for $\boldsymbol{\lambda}$ in the previous section,
\begin{align}
& \mathcal{P}(n_\mathrm{A}, n_\mathrm{B}|\gamma_i) = \mathcal{P}_{n_\mathrm{A} n_\mathrm{B}}^i, \\
& \mathcal{P}_\mathrm{A}(n_\mathrm{A}|\gamma_i) = \mathcal{P}_{n_\mathrm{A}}^i.
\end{align}
In order to estimate the left-hand side of the inequality testing the LHCS model, we represent it in the form  
\begin{align}
\boldsymbol{\lambda} \cdot \boldsymbol{\mathcal{P}} = \langle \lambda^1 \rangle + \langle \lambda^2 \rangle,
\end{align}
where
\begin{align}
\label{Eq:Lambda1}
\langle \lambda^1 \rangle = \lambda_{00}^1 \mathcal{P}_{00}^1 &+ \lambda_{01}^1 \mathcal{P}_{01}^1 \\ & + \lambda_{10}^1 \mathcal{P}_{10}^1 + \lambda_{11}^1 \mathcal{P}_{11}^1 + \lambda_0^1 \mathcal{P}_0^1,\nonumber \\
\label{Eq:Lambda2}
\langle \lambda^2 \rangle = \lambda_{00}^2 \mathcal{P}_{00}^2 &+ \lambda_{01}^2 \mathcal{P}_{01}^2 + \lambda_0^2 \mathcal{P}_0^2.
\end{align}
Utilizing identity (\ref{Eq:P02}) and introducing the notations,
\begin{align}
&\Lambda_{0,0}^1= \lambda_{00}^1 + \lambda_0^1, && \Lambda_{0,0}^2= \lambda_{00}^2 + \lambda_0^2, \\
&\Lambda_{0,1}^1=\lambda_{01}^1 + \lambda_0^1, &&\Lambda_{0,1}^2=\lambda_{01}^2 + \lambda_0^2,\\
&\Lambda_{0,2}^1= \lambda_0^1, && \Lambda_{0,2}^2= \lambda_0^2,\\
& \Lambda_{1,0}^1=\lambda_{10}^1, && \Lambda_{1,0}^2=0,\\
&\Lambda_{1,1}^1=\lambda_{11}^1, && \Lambda_{1,1}^2=0,\\
&\Lambda_{1,2}^1=0, && \Lambda_{1,2}^2=0,
\end{align}
we can rewrite these expressions as
\begin{align}
\label{Eq:Lambda}
&\langle \lambda^i \rangle 
= \sum\limits_{n_{\mathrm{A}}=0}^1\sum\limits_{n_{\mathrm{B}}=0}^2 \Lambda_{n_{\mathrm{A}} n_{\mathrm{B}}}^i\mathcal{P}_{n_{\mathrm{A}} n_{\mathrm{B}}}^i.
\end{align}
Therefore, the left-hand side of the inequality can be interpreted as the sum of  expectation values of two observables, $\lambda^1=\Lambda_{n_{\mathrm{A}} n_{\mathrm{B}}}^1$ and $\lambda^2=\Lambda_{n_{\mathrm{A}} n_{\mathrm{B}}}^2$, depending on $\gamma_1$ and $\gamma_2$, respectively. 

Let us assume that during the measurement procedure we obtain $N$ samples of pairs $(n_{\mathrm{A};j}, n_{\mathrm{B};j})$, where $j=1,\ldots,N$, for each value of the LO amplitude, $\gamma_1$ and $\gamma_2$. 
Then the estimators for $\langle\lambda^i\rangle$ are given by 
\begin{align}\label{Eq:lambda-i-est}
\langle \lambda^i \rangle \approx \theta^{i\ast}= \frac{1}{N}\sum\limits_{j=1}^{N}\Lambda_{n_{\mathrm{A};j}, n_{\mathrm{B};j}}^i.
\end{align}
The standard deviations for these estimators read
\begin{align}
\epsilon_i = \sqrt{\left\langle(\Delta\theta^{i\ast})^2\right\rangle}=
\sqrt{\frac{\big\langle \left(\lambda^i\right)^2 \big\rangle - \big\langle \lambda^i \big\rangle^2}{N} },
\end{align}
where 
\begin{align}
\label{Eq:Lambda-Sq}
&\langle (\lambda^i)^2 \rangle 
= \sum\limits_{n_{\mathrm{A}}=0}^1\sum\limits_{n_{\mathrm{B}}=0}^2 \left(\Lambda_{n_{\mathrm{A}} n_{\mathrm{B}}}^{i}\right)^2\mathcal{P}_{n_{\mathrm{A}} n_{\mathrm{B}}}^i.
\end{align}
The total error for estimation of $\boldsymbol{\lambda} \cdot \boldsymbol{\mathcal{P}}$ is given by
\begin{align}
\epsilon=\sqrt{\epsilon_1^2 + \epsilon_2^2}.
\end{align}
This error can also be estimated from the measurement data using the standard methods.

We provide an optimization procedure with respect to the coherent amplitude $\gamma_i>|\gamma_{\mathrm{min}}|$ by maximizing the relative violation of inequality (\ref{Eq:Inequalities}),
\begin{align}
V=\frac{\boldsymbol{\lambda}\cdot\boldsymbol{\mathcal{P}}- \sup_{\boldsymbol{\mathcal{A}},\alpha}\boldsymbol{\lambda}\cdot
	\boldsymbol{M}(\boldsymbol{\mathcal{A}}, \alpha)}{\epsilon}.
\end{align}
For each values of $\gamma_i$ we find vectors $\boldsymbol{\lambda}$ with the procedure involving the \textsf{QuickHull} algorithm, which is described in Sec.~\ref{Sec:Example}.
This gives us the following vectors $\boldsymbol{\lambda}=\begin{pmatrix}
\lambda_{00}^1  & \lambda_{01}^1 & \lambda_{10}^1 & \lambda_{11}^1 & \lambda_{00}^2 & \lambda_{01}^2 & \lambda_{0}^1& \lambda_{0}^2
\end{pmatrix}$ for the cases (A), (B), and (C) in Table~\ref{Tab:Violation}:

\begin{align*}
&\textrm{Case (A):}\\
&\boldsymbol{\lambda}{=}\begin{pmatrix}
0.42  & 0.43  & 0.1  & 0.17  & 0.38  & 0.46  & -0.3 & -0.39 
\end{pmatrix},\\
&\textrm{Case (B):}\\
&\boldsymbol{\lambda}{=}\begin{pmatrix}
0.48  & 0.49  & 0.1  & 0.19  & 0.32  & 0.39  & -0.35 & -0.32 
\end{pmatrix},\\
&\textrm{Case (C):}\\
&\boldsymbol{\lambda}{=}\begin{pmatrix}
0  & -0.06  & 0.11  & 0.22  & 0.52  & 0.61  & 0.15 & 0.52 
\end{pmatrix}.
\end{align*}    
        
\bibliography{biblio}	
	
\end{document}